\documentclass[10pt,letterpaper,twocolumn]{article} %% two column, final %layout

\usepackage{ol2}
\usepackage[draft]{hyperref}
\usepackage{amsmath}
\usepackage{bm,braket}

\usepackage{color}
\usepackage{ulem}
\newcommand {\be}{\begin{equation}}
\newcommand {\ee}{\end{equation}}
\newcommand {\bea}{\begin{eqnarray}}
\newcommand {\eea}{\end{eqnarray}}
\newcommand {\nn}{\nonumber}
\renewcommand{\v}[1]{\ensuremath{\mathbf{#1}}} % for vectors

\newcommand{\avg}[1]{\left< #1 \right>} % for average

\def\sm{\sigma^-}
\def\smdag{\sigma^+}
\def\f{\textbf{f}}
\def\fdag{\textbf{f}^\dagger}

\def\r1{\textbf{r}}

\begin{document}

% change spacing in align to be similar to eqnarray
\abovedisplayskip=6pt
\abovedisplayshortskip=6pt
\belowdisplayskip=6pt
\belowdisplayshortskip=6pt
\twocolumn[ %% activate for two-column option

\title{Theory of phonon-modified quantum dot photoluminescence intensity in structured photonic reservoirs }

\author{Kaushik Roy-Choudhury$^*$ and Stephen Hughes}
\affiliation{Department of Physics, Queen's University, Kingston, Ontario, Canada, K7L 3N6}
%\author{Stephen Hughes}
%\affiliation{Department of Physics, Queen's University, Kingston, Ontario, %Canada, K7L 3N6}
\email{$^*$Corresponding author: kroy@physics.queensu.ca}
%\date{\today}

\begin{abstract} 
%The broad-band frequency dependence of the local density of states of a structured photonic reservoir
The spontaneous emission rate of a quantum dot coupled to a structured photonic reservoir is determined by the  frequency dependence of its local density of photon states. Through phonon-dressing, a breakdown of Fermi's golden rule can occur for certain photonic structures whose photon decay time become comparable to the longitudinal acoustic phonon decay times. We present a polaron master equation model to calculate the photoluminescence intensity from a coherently excited quantum dot coupled to a structured photonic reservoir. We consider examples of a semiconductor microcavity and a coupled cavity waveguide and show clear photoluminescence intensity spectral features that contain unique signatures of the interplay between phonon and photon bath coupling. 
\end{abstract}

%\ocis{(270.0270) Quantum optics; 
%(350.4238) Nanophotonics and photonic crystals; (160.6000) Semiconductor %materials.} (300.6320) Spectroscopy, high-resolution; (300.6470) Spectroscopy, semiconductors

 ] %% activate for two-column option

\begin{figure}[b]
\includegraphics[height=0.78\columnwidth]{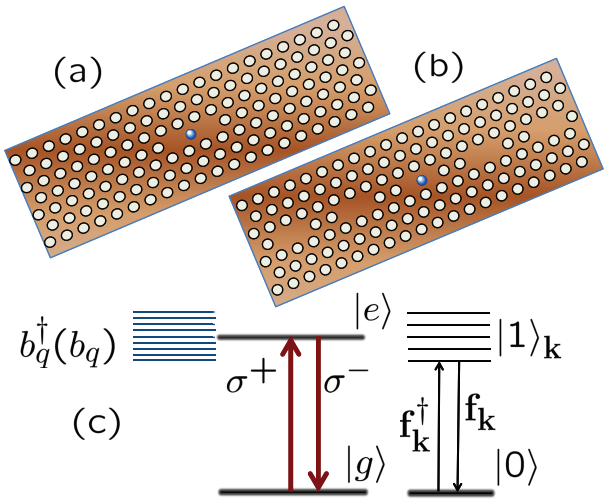}
\centering
\vspace{-0.2cm}
\caption{\label{fig1}(Color online). \footnotesize{  
Photonic microcavity (a) and a coupled-cavity
waveguide (b) using a photonic crystal platform, containing a coupled semiconductor QD. (c) Energy level diagram of a neutral QD  (electron-hole pair) interacting with a phonon bath and a photon bath. The operator ${\bf f}^\dag_{\bf k}$
($b^{\dag}_q$) creates a photon (phonon).}}
\label{fig:1}
\vspace{-0.cm}
\end{figure}

Quantum dots (QDs) behave as artificial atoms in a solid-state media and are promising for scalable quantum information processing~\cite{Kim}.
Quantum dots can also be coupled to structured photonic reservoirs like photonic crystals (PCs), which alter their light-matter interaction dynamics. However, lattice phonon interactions cause QDs to behave  differently from simple atoms~\cite{Weiler}, manifesting in various effects such as damping and frequency shifts of driven Rabi oscillations~\cite{Forstner, Ramsay, Leonard}, excitation induced dephasing of Mollow side-bands~\cite{Ulrich}, off-resonant cavity feeding~\cite{Ates,Arka} and asymmetric vacuum Rabi doublets~\cite{Milde, Ota}. Recently~\cite{Kaushik}, it has been shown
that for certain photonic reservoirs,
a 
 break-down of Fermi's golden rule can occur when  determining the spontaneous emission (SE) rate  of an embedded QD. Specifically, for reservoir decay times
that are comparable to the phonon relaxation times, 
a large bandwidth of the 
 local photon density of states (LDOS) 
 dictates how SE occurs. Thus the SE rate is not only determined by the
 LDOS at the emitter frequency~\cite{Kaushik,Valente}.

In this Letter, we develop a polaron ME approach~\cite{Kaushik} that includes both a photon reservoir and a phonon reservoir,  and explore the influence of the phonon-modified SE rate on the photoluminescence intensity (PLI). The PLI spectra is generated by recording the emission intensity of a QD, as a function of laser-exciton detuning using a coherent laser drive. This is a useful experimental technique for determining the emission properties of a QD, and recent experiments already show clear signatures of the phonon bath~\cite{Weiler, Ulrich}. Here we show theoretically how the PLI of a QD, excited by a weak coherent drive, changes in the presence of a structured photonic reservoir. We highlight several non-trivial PLI signatures that arise due to the broadband frequency dependence of the QD SE rate and perform a systematic study of these effects as a function of temperature.

Our theory  can be applied to any general LDOS medium,  but we  focus here on  microcavities and coupled cavity waveguides, as shown schematically 
 in  Fig.~\ref{fig:1}(a-b). 
%Our theory can be applied to any general LDOS function and as specific examples, we consider   cavities and coupled cavity  waveguides---see Fig.~\ref{fig:1} which also shows an energy level diagram of the QD system including both phonon and photon baths.
The semiconductor QD is modelled as a two-level exciton system (strong confinement limit) interacting with a photonic reservoir and an acoustic phonon bath~\cite{Roy2} (Fig.~\ref{fig:1}(c)). To model  the PLI, we  consider a QD that is weakly driven by a cw (continuous wave) pump laser of Rabi frequency $\eta_x$. The total Hamiltonian 
of the system, in a frame rotating at the laser frequency $\omega_L$, is~\cite{Scheel}
\begin{align}
\label{eq1}
 &H = \hbar\int d\v{r} \int_0^{\infty}d\omega\,\fdag(\r1,\omega)\f(\r1,\omega) +
\hbar\Delta_{xL} \smdag\sm \nn\\
&-\left[\smdag e^{i\omega_Lt}\int_0^{\infty}\!d\omega\,\v{d}\cdot\v{E}(\r1_d,\omega) + \text{H.c}\right] + \hbar\eta_x(\smdag+\sm) \nn\\
&+\Sigma_q\hbar \omega_q b^{\dag}_qb_q  + \smdag\sm\Sigma_q\hbar\lambda_q(b^{\dag}_q+b_q),
\end{align}
where  $\smdag$/$\sm$ are the Pauli operators of the exciton 
(electron-hole pair), $\v{d}=d\v{\hat n}_d$ is the dipole moment of the QD at spatial position ${\r1}_d$, $\Delta_{xL}=\omega_x-\omega_L$ is the exciton-laser detuning, $b_q$ ($b^{\dag}_q$) are the annihilation (creation) operators of the acoustic phonons,  and $\lambda_q$ is the exciton-phonon coupling strength; $\f$/$\fdag$ are the boson field operators of the photon reservoir, and we have used the dipole and the rotating wave approximation  in describing the interaction between the QD and photonic reservoir. The electric-field operator $\v{E}(\r1,\omega)$ is related to the 
medium  Green function $\v{G}(\r1,\r1';\omega)$~\cite{Scheel}.

The Hamiltonian $H$ is  polaron transformed as $H' \rightarrow e^P He^{-P}$ where $P = \smdag\sm \Sigma_q\frac{\lambda_q}{\omega_q} (b^{\dag}_q-b_q)$~\cite{Imamoglu}, which includes electron-phonon interaction to all orders. We assume
a weak-to-intermediate coupling between the QD and the photon bath  and 
derive a time-convolutionless~\cite{Breuer} polaron ME for the QD reduced density operator $\rho$~\cite{Kaushik} using a second-order Born approximation. The polaron ME is given by
\begin{eqnarray}
\label{eq2}
\frac{\partial \rho}{\partial t} = \frac{1}{i\hbar}[H'_S,\rho] + \mathcal{L}_{\text{phot}}(\rho) + \mathcal{L}_{\text{phon}}(\rho),
\end{eqnarray}
\noindent where $H'_S = \hbar \Delta_{xL}\smdag\sm + \hbar\eta_x\avg{B}(\smdag+\sm)$, 
%$\avg{B}$ = $\avg{B_{\pm}} =\avg{ \exp[\pm\Sigma_q\frac{\lambda_q}{\omega_q}(b_q-b^{\dag}_q)]}$
$\braket{B} = \exp[-\frac{1}{2}\phi(0)]$ is the thermal average of the  phonon bath displacement operator~\cite{Imamoglu}, and 
$\phi(t) = \int_0^{\infty} d\omega\frac{J_{\rm pn}
(\omega)}{\omega^2}[\coth(\hbar\omega/2k_BT)\cos(\omega t)- i\sin(\omega t)]$ is the independent-boson model phase with $J_{\rm pn}(\omega)$  the phonon spectral function~\cite{Roy2}.
For convenience, a polaron shift $\Delta_P= \int_0^{\infty}d\omega\frac{J_{\rm pn}(\omega)}{\omega}$ is implicitly included in the definition of $\Delta_{xL}$. A trace over the phonon~\cite{Roy2} and photon variables~\cite{Tanas} is performed in deriving Eq.~\eqref{eq2} where the phonon and photon reservoirs are assumed to be in thermal equilibrium and statistically 
independent~\cite{Carmichael}.

%For optical photon energies, the temperature of the photon bath is assumed %to be 0 K. 

The incoherent interaction of the QD with the photon and the phonon reservoirs are described by the superoperator terms $\mathcal{L}_{\text{phot}}$ and $\mathcal{L}_{\text{phon}}$, respectively; $\mathcal{L}_{\text{phon}}$ is not influenced by the photonic reservoir and in the limit of small pump rates can be approximated as $\mathcal{L}_{\text{phon}}= \Gamma^{\sigma^{+}}{L}[\sigma^+]+\Gamma^{\sigma^{-}}{L}[\sigma^-] - \gamma_{\rm cd}(\smdag\rho\smdag+\sm\rho\sm)$, where ${L}[O]
=\frac{1}{2}(2O\rho O^\dagger -O^\dagger O \rho-\rho O^\dagger O)$. The  phonon scattering terms, ${\Gamma^{\sigma^{+/-}}}=2\avg{B}^2\eta_x^2 \text{Re}[\int_0^{\infty}d\tau e^{\mp i \Delta_{xL}\tau}(e^{\phi(\tau)} - 1)]$ represent incoherent excitation and radiative decay of the QD, respectively~\cite{Roy3}, and the cross-dephasing term $\gamma_{\rm cd} = 2\avg{B}^2\eta_x^2 \text{Re}[\int_0^{\infty}d\tau {\rm cos}(\Delta_{xL}\tau)(1-e^{-\phi(\tau)})]$ influences the spectral lineshape through dephasing of off-diagonal density matrix elements~\cite{Ulhaq}. The interaction with the photonic reservoir is however modified by phonons, and is given by $\mathcal{L}_{\text{phot}}(\rho) = \int_0^t d\tau \int_0^{\infty} d\omega\,  {J_{\text{ph}}(\omega)}[-C_{\text{pn}}(\tau)\smdag\sm(-\tau)e^{i\Delta_L\tau}\rho + C^*_{\text{pn}}(\tau)\sm\rho\smdag(-\tau)e^{-i\Delta_L\tau}+C_{\text{pn}}(\tau)\sm(-\tau)\rho\smdag e^{i\Delta_L\tau} - C^*_{\text{pn}}(\tau)\rho\smdag(-\tau)\sm e^{-i\Delta_L\tau}]$,~\cite{Kaushik} where $\Delta_L = \omega_L-\omega$, $J_{\text{ph}}(\omega)= \frac{\v{d}\cdot \text{Im}[\v{G}(\r1_d,\r1_d;\omega)]\cdot \v{d}}{\pi\hbar\epsilon_0}$ is the photon reservoir spectral function, while  $C_{\text{pn}}(\tau)=e^{[\phi(\tau)-\phi(0)]}$ is the {\it phonon}  correlation function. The time-dependent operator $\sigma^{\pm}(-\tau) = e^{-iH'_S\tau/\hbar} \sigma^{\pm}e^{iH'_S\tau/\hbar}$ results in pump field dependent scattering terms; but for weak drives, such dependence is negligible and $\text{Re}[\mathcal{L}_{\text{ph}}]$ gives rise to phonon-modified spontaneous emission decay ($\tilde{\gamma} L[\sigma^-]$), where the phonon-modified SE decay rate  is~\cite{Kaushik,Nazir2}
\begin{align}
\tilde{\gamma} = 2\int_0^{\infty}\text{Re}[C_{\text{pn}}(\tau)J_{\text{ph}}(\tau)]d\tau,
\label{eq:SE}
\end{align}
 and $J_{\text{ph}}(\tau) = \int_0^{\infty} d\omega J_{\text{ph}}(\omega) e^{i(\omega_L-\omega)\tau}$ is the photon correlation function.
%\com{Here and below is too similar to other paper - rephrase everything!!!}
 As shown elsewhere~\cite{Kaushik}, this phonon-modified SE rate contains contribution from the broadband photonic LDOS sampled by the bandwidth of the phonon bath, which is in contradiction with the well known Fermi's golden rule for SE decay. A measurement of SE rate as a function of the photonic LDOS frequency can  demonstrate the influence of  non-local LDOS contributions ~\cite{Kaushik}. Such measurements are however difficult to obtain experimentally. Alternatively, a PLI measurement from a coherently excited QD can
be used to probe the non-local frequency dependence of SE rate. Such measurements have been used to measure the  phonon side-bands  from a single excited QD, but without any photon reservoir coupling~\cite{Ulhaq}.
Thus it is of practical relevance to study the effects of  photon reservoir coupling.

The QD PLI ($I_{\rm x}$) is  proportional to exciton population ${\rm n_x} = \avg{\smdag\sm}={\rm Tr}(\smdag\sm\rho)$, where ${\rm Tr}$ denotes the trace. For the current problem, an analytical form of $n_{\rm x}$ is  derived using the Bloch equations~\cite{Ulhaq}, yielding 

\begin{align}
\label{eq5}
{\rm n_x} = \frac{1}{2}\left[1+\frac{\Gamma^{\smdag} - \Gamma^{\sm} -\tilde\gamma}{\Gamma^{\smdag} + \Gamma^{\sm} 
+\tilde\gamma+\frac{4\avg{B}^2\eta^2_x(\Gamma_{\text{pol}}
+\gamma_{\rm cd})}{\Gamma^2_{\text{pol}}+\Delta^2_{xL}-\gamma^2_{\rm cd}}}
\right],
\end{align}
where $\Gamma_{\text{pol}} = \frac{1}{2}(\Gamma^{\smdag} + \Gamma^{\sm} +\tilde\gamma + \gamma')$. A temperature dependent pure dephasing term of the form $\gamma' = 3+0.95(T-1) \mu$eV ~\cite{Ota,Borri} is also included in the ME (Eq.~\eqref{eq2}) for calculation of the PLI.

To observe the effects of  reservoir coupling on the PLI, we choose a Lorentzian single mode PC cavity (Fig.~\ref{fig:1}(a)) and a PC coupled-cavity waveguide (Fig.~\ref{fig:1}(b)), both in the weak coupling regime. The PC defect cavities (Fig.~\ref{fig:1}(a)) are important for investigating fundamental aspects of cavity-QED in solid-state~\cite{Yoshie}, with potential applications for quantum information processing~\cite{Kim} and low power optoelectronics~\cite{Bose}. Photonic crystal waveguides (Fig.\ref{fig:1}(b)) are useful for slow-light propagation~\cite{Notomi1} and for manipulating the emission properties of embedded QDs  for on-chip single photon emission~\cite{Rao1, Shields}. 
The bath function for a single cavity mode is $
J_{\text{ph}}(\omega) 
=g^2\frac{1}{\pi}\frac{\frac{\kappa}{2}}{(\omega-\omega_c)^2+(\frac{\kappa}{2})^2}$, where $g$
is the  QD-cavity coupling rate, and $\kappa$ is the cavity decay rate. 
For the
 PC coupled cavity waveguide~\cite{Yariv}, the photon bath function is evaluated using an analytical tight-binding technique, where nearest neighbor coupling is assumed between adjacent cavities of mode volume $V_{\rm eff}$ ~\cite{Fussell1}, $J_{\text{ph}}(\omega) 
\!=\!\frac{-{d}^2\omega}{2\hbar\epsilon_0n_b^2 V_{\rm eff}} \frac{1}{\pi}
{\rm Im}\left[ \frac{1}{\sqrt{(\omega-\tilde\omega_u)(\omega-\tilde\omega_l^*)}}\right]$, where $\tilde\omega_{u,l}=\omega_{u,l}\pm i\kappa_{u,l}$\cite{Fussell1}. The real part ($\omega_{u,l}$) represents the band-edge frequencies of the waveguide (see Fig.~\ref{fig3}(a)), the imaginary part $\kappa_{u,l}$ represents damping and $n_b$ is the refractive index of the  dielectric. 
For our phonon calculations we use parameters describing InAs QDs~\cite{Ulhaq}, where longitudinal acoustic  phonon interaction, resulting from deformation potential coupling dominates and $J_{\rm pn}(\omega) = \alpha_p\omega^3 \exp[-\frac{\omega^2}{2\omega_{ b}^2}]$. The phonon cutoff frequency due to spatial charge confinement is $\omega_{ b} = 1$ meV, and exciton-phonon coupling strength  $\alpha_{ p}/(2\pi)^2 = 0.06 \rm\, ps^2$ \cite{Weiler}. 
The QD dipole moment is taken to be
$d = 50$ Debye.

%We note that the imaginary part of $\mathcal{L}_{\text{ph}}$ yield small Lamb shifts~\cite{Roy3}, which are safely ignored in the  calculations below.

% something strange here but this fixes it!!
\vspace{-0.01cm}

\begin{figure}[t]
\vspace{0.cm}
\includegraphics[width=0.99\columnwidth]{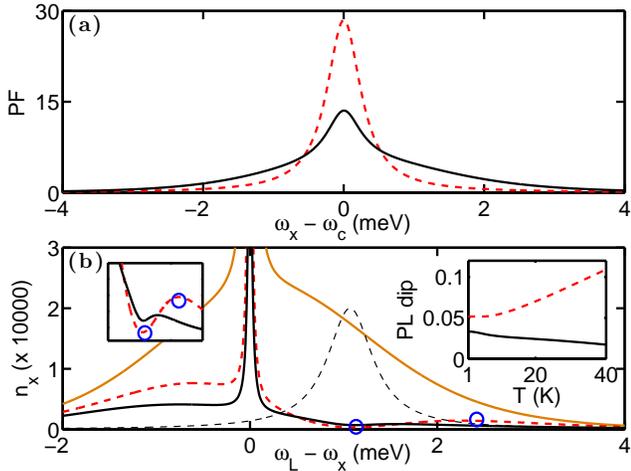}
\vspace{-0.cm}
\caption{\label{fig1}(Color online). \footnotesize{{\bf Cavity}: (a) Purcell factor at T = 40 K for cavity with (solid) and without phonons (dashed). The cavity decay rate, $\kappa$ = 0.6 meV which corresponds to a Q of 2300 at $\omega_c/2\pi$ = 1440 meV. (b) QD excitation ${\rm n_x}$ ($\propto I_{\rm x}$), when the cavity is aligned to the right of the QD at T = 40 K. The photonic reservoir line-shape (dark dashed line) is plotted for reference. %and the parameters are same as in Fig.~\ref{fig:2}(a). 
The solid light line is the excitation in the absence of photonic reservoir coupling. The dark solid (light dashed) line plots ${\rm n_x}$ in the presence of the photonic reservoir (cavity structure), when phonons do (do not) influence the SE rate, $\tilde\gamma$ ($\gamma$). Left inset is a magnified view of the PLI dip in the presence of structured photon reservoirs. Right inset shows the variation of the PLI dip with temperature. Solid (dashed) line shows the intensity dip with (without) phonon modification to the SE. }}
\label{fig:2}
\end{figure}

\begin{figure}[th!]
\vspace{0.cm}
\includegraphics[width=0.98\columnwidth]{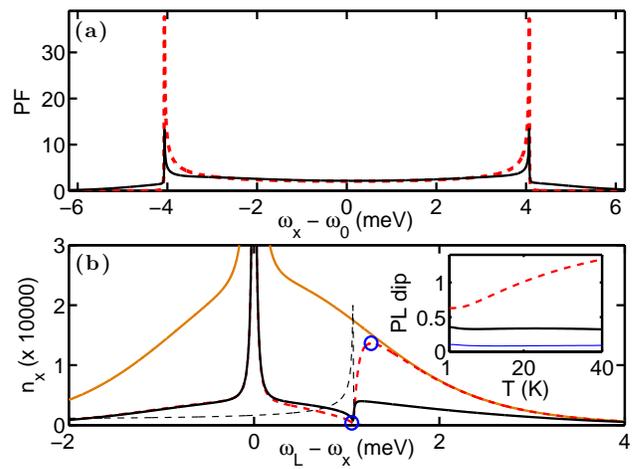}
\vspace{-0.cm}
\caption{\label{fig3}(color online). \footnotesize{ {\bf Waveguide}: (a) Purcell factor at T = 40 K for a coupled-cavity waveguide with (solid) and without phonons (dashed), plotted as a function of detuning from band-center $\omega_0$. The waveguide calculation uses parameters from~\cite{Kuramochi}.
(b) QD excitation ${\rm n_x}$ ($\propto I_{\rm x}$) when the waveguide upper mode-edge is aligned to the right of the QD at T = 40 K. 
%The parameters for the waveguide are same as in Fig.~\ref{fig3}(a). 
The curves represent the same quantities as in Fig.~\ref{fig:2}(b). The solid thin line in the lower inset shows the PLI dip with  a symmetric Lorentzian cavity.}}
\end{figure}
%The waveguide calculations uses parameters from Ref.~\onlinecite{Kuramochi}.

The PLI ($\propto {\rm n_x}$) from a QD  coupled to a cavity and a waveguide is plotted in Fig.~\ref{fig:2}(b)  and  Fig.~\ref{fig3}(b), respectively. To help explain the spectral profile, we have also reproduced the Purcell factors (PF) (top panels, Fig. \ref{fig:2}(a), \ref{fig3}(a)) for these two representative LDOS profiles~\cite{Kaushik}. The PF is defined as ${\rm PF} =\tilde\gamma/\gamma_b$, where $\gamma_b$ represents the decay rate in the background slab material. The solid and dashed lines are PFs with and without phonons, respectively, and the PF without phonon interactions is  $\gamma/\gamma_b$, where $\gamma = 2\int_0^{\infty}\text{Re}[J_{\text{ph}}(\tau)]d\tau$. In the  regions where the contribution from the local photonic LDOS (cavity peak, waveguide mode-edge) is strong, phonons clearly reduce the PF. In contrast, in regions where the photonic LDOS is weak, then phonons enhance the SE rate through non-local contributions of the photonic LDOS. 

For our PLI  calculations (Fig.~(\ref{fig:2}, \ref{fig3}) (b)), the QD is held at a fixed detuning from the photonic LDOS and is excited with a weak drive laser ($\eta_x = 0.4 \mu$eV), where the laser frequency is varied
to obtain the PLI as a function of laser-exciton detuning.  The light solid line in Fig.~(\ref{fig:2}, \ref{fig3}) (b) is the PLI in the absence of any structured reservoirs ($\gamma_b = 1\mu$eV) and $\Delta_{xL} = 0$ meV marks the position of the zero phonon line (ZPL). The cavity (Fig.~\ref{fig:2}(b)) and the waveguide upper mode-edge (Fig.~\ref{fig3}(b)) is located approximately 1\,meV to the right of the ZPL (dark dashed line). 
In the absence of a structured photonic reservoir, the PLI is enhanced around this region at low temperatures~\cite{Weiler}.  The phonon-assisted incoherent excitation process $\Gamma^{\smdag}$ increases with laser detuning $\Delta_{Lx}$ due to higher phonon  emission probability and the phonon-induced decay rate $\Gamma^{\sm}$ reduces~\cite{Roy3}. This leads to larger population excitation ${\rm n_ x}$ (Eq.\eqref{eq5}) and hence stronger QD emission on the blueside of the ZPL.  However, the presence of the structured reservoirs reduces the PLI (light dashed line). For weak driving, the SE rate $\gamma$ samples the photonic LDOS at the drive laser frequency ($\omega_L$)
when the Markov approximation is valid~\cite{John2,Scheel,Xue}. 
Hence $\gamma$, following the photonic LDOS, is enhanced in this region. Thus the PLI shows a dip (Fig.~\ref{fig:2}(b) left inset and Fig.~\ref{fig3}(b)). When we include a  phonon-modified SE rate ($\tilde\gamma$), the size of this dip reduces (dark solid line). It should be noted that in the case of a cavity, $\tilde\gamma$ can also be estimated using phonon-mediated cavity scattering rates~\cite{Roy3}, in the high $Q$ limit approximation~\cite{Kaushik}. However for the cavity parameters used, there is negligible difference in the PLI  profiles above. However, for smaller  Q cavities (i.e., $Q< 1000$), these two ME techniques result in substantially different SE rates $\tilde\gamma$, though the
 PLI dip is  smaller. Thus cavities with Q factors greater than a few thousand
 are better to capture the non-local LDOS effects through PLI  measurements; here any departure from a symmetric Lorentzian LDOS would require the full reservoir calculation of $\tilde\gamma$ (Eq.\eqref{eq:SE}) to understand the PLI profiles. 
%An example of this is shown below for the waveguide.

In the case of a cavity,
the broadband frequency dependence of SE can be observed from the line-shape of the PLI dip in Fig.~(\ref{fig:2}(b) left inset). Furthermore the dependence of the PLI dip on temperature bears a clearer signature of this effect. The PLI dip is estimated from the difference in intensity between the points marked by the circles in Fig.~\ref{fig:2}(b) and \ref{fig3}(b). The circles mark the PLI dip and the highest intensity on the right of the dip. When the SE is not influenced by phonons, the PLI dip increases as a function of temperature as the  emission intensity of the phonon side-band increases (Fig.~\ref{fig:2}(b), \ref{fig3}(b)) inset, dashed line).  When phonon effects are included in the SE  rate, then the size of this PL dip reduces (Fig.~\ref{fig:2}, \ref{fig3} (b)) inset, thick solid line). This is caused by a phonon-induced reduction of SE close to peak LDOS and an enhancement away from the peak LDOS (dark solid line, Fig.\ref{fig3} (a)). Such behavior is more discernable in the case of a waveguide (Fig.~\ref{fig3}(b) inset, thick solid line), due to the asymmetric nature of its mode-edge LDOS~\cite{Kaushik}. Note that a symmetric Lorentzian cavity of similar linewidth (Fig.~\ref{fig3} (b) inset, thin solid line) produces a much smaller PLI dip. 
%This clearly shows the broad-band frequency dependence of SE rates in structured reservoirs.  

In conclusion, we have presented a theoretical study to describe PLI as a function of laser-exciton detuning, from a coherently excited QD coupled to a structured photonic reservoir in the presence of electron-phonon coupling. Using the examples of a cavity and a slow-light waveguide, we have demonstrated how the frequency dependence of the phonon-modified SE rate influences the  PLI, causing clear spectral signatures that should be observable in related experiments. Our theory can be used to model the QD PLI in a wide range of photon reservoirs.

%We have 
%exemplified our theory by studying the modified SE rate from both simple %cavity structures and CROWs.

This work was supported by the Natural Sciences and Engineering Research Council of Canada.

\newpage

\section*{Informational Fifth Page}
In this section, we provide full versions of citations to
assist reviewers and editors.


\begin{thebibliography}{20}



\bibitem{Kim} H. Kim, R. Bose, T. Shen, G. Solomon and E. Waks, \textrm{A quantum logic gate between a solid-state quantum bit and a photon.}  Nature Photon. {\bf 7,} 373 (2013).

\bibitem{Weiler} S. Weiler, A. Ulhaq, S. M. Ulrich, D. Richter, M. Jetter, P. Michler, C. Roy, and S. Hughes, \textrm{Phonon-assisted incoherent excitation of a quantum dot and its emission properties.} { Phys. Rev. B} {\bf 86,} 241304(R) (2012).

%\bibitem{Axt2} B. Krummheuer, V. M. Axt and T. Kuhn, \textrm{Theory of pure dephasing and the resulting absorption line shape in semiconductor quantum dots.} { Phys. Rev. B} {\bf 65,} 195313 (2002).

\bibitem{Forstner} J. F\"{o}rstner, C. Weber, J. Danckwerts and A. Knorr, \textrm{Phonon-Assisted Damping of Rabi Oscillations in Semiconductor Quantum Dots.} { Phys. Rev. Lett.} {\bf 91,} 127401 (2003).

\bibitem{Ramsay} A. J. Ramsay, T. M. Godden, S. J. Boyle, E. M. Gauger, A. Nazir, B. W. Lovett, A. M. Fox, and M. S. Skolnick, \textrm{Phonon-Induced Rabi-Frequency Renormalization of Optically Driven Single InGaAs/GaAs Quantum Dots.} { Phys. Rev. Lett.} {\bf 105,} 177402 (2010).

\bibitem{Leonard} L. Monniello, C. Tonin, R. Hostein, A. Lemaitre, A. Martinez, V. Voliotis, and R. Grousson, \textrm{Excitation-Induced Dephasing in a Resonantly Driven InAs/GaAs Quantum Dot.} {Phys. Rev. Lett.} {\bf 111,} 026403 (2013).

\bibitem{Ulrich} S. M. Ulrich, S. Ates, S. Reitzenstein, A. L\"{o}ffler, A. Forchel, and P. Michler, \textrm{Dephasing of Triplet-Sideband Optical Emission of a Resonantly Driven InAs/GaAs Quantum Dot inside a Microcavity.} {Phys. Rev. Lett.} {\bf 106,} 247402 (2011).

\bibitem{Ates} S. Ates, S. M. Ulrich, A. Ulhaq, S. Reitzenstein, A. L\"{o}ffler, S. H\"{o}fling, A. Forchel and P. Michler, \textrm{Non-resonant dot–cavity coupling and its potential for resonant single-quantum-dot spectroscopy.} Nature Photon. {\bf 3}, 724 (2009).

\bibitem{Arka} A. Majumdar, E. D. Kim, Y. Gong, M. Bajcsy, and J. Vu\u{c}kovi\'{c}, \textrm{Phonon mediated off-resonant quantum dot–cavity coupling under resonant excitation of the quantum dot.} { Phys. Rev. B} {\bf 84,} 085309 (2011).

\bibitem{Milde} F. Milde, A. Knorr and S. Hughes \textrm{Role of electron-phonon scattering on the vacuum Rabi splitting of a single-quantum dot and a photonic crystal nanocavity.} {Phys. Rev. B} {\bf 78,} 035330 (2008).

\bibitem{Ota} Y. Ota, S. Iwamoto, N. Kumagai \& Y. Arakawa, \textrm{Impact of electron-phonon interactions on quantum-dot cavity quantum electrodynamics.} arXiv:0908.0788 (2009).

\bibitem{Kaushik} K. Roy-Choudhury \& S. Hughes, \textrm{Phonon-modified spontaneous emission from single quantum dots in a structured photonic medium} arXiv:1406.3649 (2014).

\bibitem{Valente} D. Valente, J. Suffczy\'{n}ski, T. Jakubczyk, A. Dousse, A. Lema\^{i}tre, I. Sagnes, Lo\"{i}c Lanco, P. Voisin, A. Auff\'{e}ves, and P. Senellart, \textrm{Frequency cavity pulling induced by a single semiconductor quantum dot.} {Phys. Rev. B} {\bf 89,} 041302(R) (2014).

\bibitem{Roy2} C. Roy and S. Hughes, \textrm{Phonon-Dressed Mollow Triplet in the Regime of Cavity Quantum Electrodynamics: Excitation-Induced Dephasing and Nonperturbative Cavity Feeding Effects.} { Phys. Rev. Lett.} {\bf 106,} 247403 (2011).

\bibitem{Scheel}  S. Scheel,  L. Kn\"{o}ll and D.-G. Welsch, \textrm{Spontaneous decay of an excited atom in an absorbing dielectric.} { Phys. Rev. A} {\bf 60,} 4094 (1999).

\bibitem{Imamoglu} I. Wilson-Rae and  A. Imamo\u{g}lu,  \textrm{Quantum dot cavity-QED in the presence of strong electron-phonon interactions.} { Phys. Rev. B} {\bf 65,} 235311 (2002).

\bibitem{Breuer} H.-P. Breuer  and  F.  Petruccione,  {\it The Theory Of Open Quantum Systems} (Oxford Univ. Press, Oxford, 2002).


\bibitem{Tanas} A. Kowalewska-Kud\l ask and R. Tana\'{s},  \textrm{Generalized master equation for a two-level atom in a strong field and tailored reservoirs.} {  J. Mod. Opt.} {\bf 48,} 347 (2001).

\bibitem{Carmichael} H. J. Carmichael, {\it Statistical Methods in Quantum Optics 1: Master Equations and Fokker-Planck Equations} (Springer, 2003).



\bibitem{Roy3} C. Roy  and S. Hughes,  \textrm{Influence of Electron$-$Acoustic Phonon Scattering on Intensity Power Broadening in a Coherently Driven Quantum Dot$-$Cavity System.} { Phys. Rev. X} {\bf 1,} 021009 (2011).

\bibitem{Ulhaq}  A. Ulhaq, S. Weiler, C. Roy, S. M. Ulrich, M. Jetter, S. Hughes, and P. Michler, \textrm{Detuning-dependent Mollow triplet of a coherently-driven single quantum dot.} { Optics Express} {\bf 21,} 4382 (2013).

\bibitem{Nazir2} D. P. S. McCutcheon and A. Nazir, \textrm{Model of the Optical Emission of a Driven Semiconductor Quantum Dot: Phonon-Enhanced Coherent Scattering and Off-Resonant Sideband Narrowing.} Phys. Rev. Lett. {\bf 110,} 217401 (2013).

\bibitem{Borri} P. Borri, W. Langbein, S. Schneider, U. Woggon, R. L. Sellin, D. Ouyang, and D. Bimberg, \textrm{Ultralong Dephasing Time in InGaAs Quantum Dots.}  { Phys. Rev. Lett.} {\bf 87,} 157401 (2001).
%\bibitem{Ge2013} 
%R. C. Ge, C. Van Vlack, P. Yao, J. F. Young, S. Hughes, 
%\textrm{Accessing quantum nanoplasmonics in a hybrid quantum-dot metal nanosystem: Mollow triplet of a quantum dot near a metal nanoparticle.}  
%{Phys. Rev. B} {\bf 87,} 205425 (2013).
%
%\bibitem{Nazir} D. P. S.  McCutcheon and  A. Nazir, \textrm{Model of the Optical Emission of a Driven Semiconductor Quantum Dot: Phonon-Enhanced Coherent Scattering and Off-Resonant Sideband Narrowing.} { Phys. Rev. Lett.} {\bf 110,} 217401 (2013).
%
%
%\bibitem{Roy1} C. Roy  and S. John, \textrm{Microscopic theory of multiple-phonon-mediated dephasing and relaxation of quantum dots near a photonic band gap.} { Phys. Rev. A} {\bf 81,} 023817 (2010).



\bibitem{Yoshie} T. Yoshie, A. Scherer, J. Hendrickson, G. Khitrova, H. M. Gibbs, G. Rupper, C. Ell, O. B. Shchekin and D. G. Deppe, \textrm{Vacuum Rabi splitting with a single quantum dot in a photonic crystal nanocavity.} {\it Nature} {\bf 432}, 200-203, (2004).

\bibitem{Bose} R. Bose, D. Sridharan, G. S. Solomon,  and E. Waks, \textrm{Low-Photon-Number Optical Switching with a Single Quantum Dot Coupled to a Photonic Crystal Cavity.} Phys. Rev. Lett. {\bf 108}, 227402 (2012).




\bibitem{Notomi1} M. Notomi, K. Yamada, A. Shinya, J. Takahashi, C. Takahashi, and I. Yokohama, \textrm{Extremely Large Group-Velocity Dispersion of Line-Defect Waveguides in Photonic Crystal Slabs.} {Phys. Rev. Lett.} {\bf 87,} 253902 (2001).


\bibitem{Rao1} V. S. C. Manga Rao and S. Hughes, \textrm{Single quantum-dot Purcell factor and $\beta$ factor in a photonic crystal waveguide.} { Phys. Rev. B} {\bf 75,} 205437 (2007).


 \bibitem{Shields} A. Schwagmann, S. Kalliakos, I. Farrer, J. P. Griffiths, G. A. C. Jones, D. A. Ritchie and A. J. Shields, \textrm{On-chip single photon emission from an integrated semiconductor quantum dot into a photonic crystal waveguide.} { Appl. Phys. Lett.} {\bf 99,} 261108 (2011). 

\bibitem{Yariv}  A. Yariv,  Y. Xu, R. K. Lee and A. Scherer, \textrm{Coupled-resonator optical waveguide: a proposal and analysis.} { Opt. Lett.} {\bf 24,} 711 (1999).



\bibitem{Fussell1} D. P. Fussell and M. M. Dignam, \textrm{Quantum-dot photon dynamics in a coupled-cavity waveguide: Observing band-edge quantum optics.} { Phys. Rev. A} {\bf 76,} 053801 (2007).

\bibitem{Kuramochi} E. Kuramochi, M. Notomi, S. Mitsugi, A. Shinya, T. Tanabe, and T. Watanabe, \textrm{Ultrahigh-Q photonic crystal nanocavities realized by the local width modulation of a line defect.} { Appl. Phys. Lett.} {\bf 88,} 041112 (2006).

%\bibitem{Fussell2} D. P. Fussell,  S. Hughes and  M. M. Dignam,  \textrm{Influence of fabrication disorder on the optical properties of coupled-cavity photonic crystal waveguides.} { Phys. Rev. B} {\bf 78,} 144201 (2008).
%
%\bibitem{Notomi2} M. Notomi, E. Kuramochi and T. Tanabe \textrm{Large-scale arrays of ultrahigh-Q coupled nanocavities.}  Nature Photon. {\bf 2,} 741 (2008).






\bibitem{John2} M. Florescu and S. John, \textrm{Single-atom switching in photonic crystals.} { Phys. Rev. A} {\bf 64,} 033801 (2001).

\bibitem{Xue} Xue-Wen Chen, Vahid Sandoghdar, and Mario Agio, \textrm{Coherent Interaction of Light with a Metallic Structure Coupled to a Single Quantum Emitter: From Superabsorption to Cloaking.} Phys. Rev. Lett. {\bf 110}, 153605.

\end{thebibliography}
\end{document}